\documentclass[12pt]{article}
\usepackage{color}
\usepackage{times}
\usepackage{graphicx,amsmath,amssymb}
\newcounter{lastnote}
\newenvironment{scilastnote}{%
\setcounter{lastnote}{\value{enumiv}}%
\addtocounter{lastnote}{+1}%
\begin{list}%
{\arabic{lastnote}.}
{\setlength{\leftmargin}{.22in}}
{\setlength{\labelsep}{.5em}}}
{\end{list}}

\title{Multiple Energy Scales at a Quantum Critical Point}

\author{P.~Gegenwart,$^{1,\dag,\ast} $ T.~ Westerkamp,$^{1}$ C.~Krellner,$^{1}$
Y.~Tokiwa,$^{1,\ddag}$ S.~Paschen,$^{1,\S}$\\ C.~Geibel,$^{1}$ F.~Steglich,$^{1}$
E.~Abrahams,$^{2}$ Q.~Si$^{3,\ast}$\\
\\
\normalsize{$^{1}$Max Planck Institute for Chemical Physics of
Solids, D - 01187~Dresden, Germany}\\
\normalsize{$^{2}$Center for Materials Theory, Department of Physics and
Astronomy, Rutgers University,}\\
\normalsize{Piscataway, New Jersey 08855, USA}\\
\normalsize{$^{3}$ Department of Physics \& Astronomy, Rice University,
Houston, TX 77005, USA}\\
\\
\normalsize{$^\ast$To whom correspondence should be addressed; E-mails:}\\
\normalsize{pgegenw@gwdg.de; qmsi@rice.edu.}\\
\\
\normalsize{$^{\dag}$Present address:
First Physics Institute, University of Goettingen, 37077 Goettingen, Germany, }\\
\normalsize{$^{\ddag}$Present address: Los Alamos National Laboratory, Los Alamos, New Mexico 87545, USA}\\
\normalsize{$^{\S}$Present address: Institute of Solid State Physics, Vienna University of Technology,}\\
\normalsize{ 1040 Vienna, Austria}
 }

\date{}
\begin{document}
\newcommand{\gc}{\color{green}}
\baselineskip24pt \maketitle

% Place your abstract within the special {sciabstract} environment.
\begin{abstract}
We report thermodynamic measurements in a magnetic-field-driven
quantum critical point of a heavy fermion metal, YbRh$_2$Si$_2$. The
data provide evidence for an energy scale in the equilibrium
excitation spectrum, that is in addition to the one expected from
the slow fluctuations of the order parameter. Both energy scales
approach zero as the quantum critical point is reached, thereby
providing evidence for a new class of quantum criticality.
\end{abstract}
% In setting up this template for *Science* papers, we've used both
% the \section* command and the \paragraph* command for topical
% divisions.  Which you use will of course depend on the type of paper
% you're writing.  Review Articles tend to have displayed headings, for
% which \section* is more appropriate; Research Articles, when they have
% formal topical divisions at all, tend to signal them with bold text
% that runs into the paragraph, for which \paragraph* is the right
% choice.  Either way, use the asterisk (*) modifier, as shown, to
% suppress numbering.
\newpage
Quantum criticality encodes the strong fluctuations of matter
undergoing a second order phase transition at zero temperature. It
underlies the unusual properties observed in a host of quantum
materials. A basic question that remains unsettled concerns its
proper theoretical description, which is challenging because the
fluctuations are both collective and quantum mechanical. One class
of theory, based on the traditional formulation of classical
critical phenomena~\cite{Wilson}, considers the fluctuations of a
classical variable -- Laudau's order parameter --
in both spatial and temporal dimensions~\cite{Hertz,Millis,Moriya,Mathur}.
The slowing down of the order-parameter fluctuations accompanies the
divergence of a spatial correlation length;
at each value of the
tuning parameter, the equilibrium many-body spectrum
contains
a single
excitation
energy
scale,
which
vanishes at the quantum critical point (QCP)~\cite{Sachdev}.
An unconventional class
of theory~\cite{Coleman,Si,Senthil}, by contrast, is inherently
quantum
mechanical; it explicitly invokes quantum entanglement effects, which
are manifested through vanishing energy scale(s) that are in
addition to
the one
associated with the slowing down of
order-parameter fluctuations.
The nature of quantum criticality can therefore be experimentally
elucidated by determining whether a single or multiple energy scale(s)
vanish as the QCP is reached.

%%In this paper, w
We consider the heavy-fermion metal YbRh$_2$Si$_2$ (YRS),
%%.We
and
show that multiple
energy scales
vanish
as its QCP is approached
and, in addition, suggest that
critical electronic modes co-exist with the slow fluctuations of the
magnetic order parameter. A direct way to probe the intrinsic energy
scales in the equilibrium spectrum near a QCP is to measure
thermodynamic properties.
Another approach is to
measure the fluctuation spectrum in equilibrium, for example by
inelastic neutron scattering experiments. Such equilibrium methods
are in contrast to transport experiments, which are influenced by
electronic relaxational properties, especially for anisotropic and
multi-band systems.

%%Since to extract
As extraction of
critical energy
scales
requires measurements
through fine steps of the control parameter, which is nearly impossible
for inelastic neutron scattering,  we report here measurements
of thermodynamic properties of YRS across its magnetic QCP.
%%To our knowledge, no prior
%%thermodynamic experiment has revealed more than one vanishing energy
%%scale
%%on approach to
%%any QCP.

We
%%have addressed this fundamental issue using
choose to work with
the tetragonal heavy
fermion compound YRS
%%, because it is advantageous to do so with
as it presents
a clean and stoichiometric material that is well
characterized~\cite{Trovarelli}. In the absence of an external
magnetic
field, YRS shows very weak antiferromagnetic (AF) order at
$T_{N}=70$ mK with an ordered moment of only $\sim
\!\!10^{-3}\mu_B$/Yb \cite{Ishidausr}. A small magnetic field ($
H_{\perp c}\approx 0.06$ T, for the field applied within the easy
$ab$ plane, and $H_{\parallel c} = 0.66$ T, along the hard $c$ axis)
suppresses the transition temperature and accesses the
QCP~\cite{Gegenwart02}. The ability of using such a small magnetic
field to access the QCP makes YRS
%%uniquely
suited for our purpose;
the determination of energy scales requires scanning across the
phase transition, and an external magnetic field can be tuned with
relative ease and continuously.
Hall effect measurements~\cite{Paschen}
%% have been carried out in
on
YRS
%%; these
have shown a large and rapid crossover in the Hall
constant at a temperature-dependent magnetic field away from
the antiferromagnetic transition.
In the zero-temperature limit, this crossover extrapolates to a jump
across the QCP, which has been interpreted as a large change of
the Fermi surface volume.
This
represents yet another advantage of measuring the thermodynamic
properties
in YRS, as they can be compared with their
transport counterparts.

We
%%have
measured the isothermal linear magnetostriction $\partial
\ln L /\partial H$, where $L$ is the length along the $[110]$
direction within the tetragonal $ab$ plane and the magnetic field
$H$ is applied along the same direction ($H\!\perp\!c$). Fig.~1
shows the magnetostriction as a function of the magnetic field, at
temperatures ranging from 0.02~K to 0.8~K. For temperatures below
0.075~K, a clear discontinuity is observed when suppressing the AF
order by a critical magnetic field. At $T>0.075$~K, it is seen that,
for small magnetic field,
the isothermal magnetostriction linearly depends on
the magnetic field, as is the case in typical metals~\cite{Chadrasekhar}.
Beyond a crossover field, however, there is a
change to a high
field region with a different slope.
The crossover field decreases as the temperature is reduced.

To understand this
crossover, we compare it with the field-dependent isothermal
behavior of other thermodynamic and transport quantities.
Fig~2{\bf A} illustrates the
%%remarkable
similarity of the
crossover in
the magnetostriction with that seen in the field-dependent
isothermal Hall resistivity $\rho_H$ (measured with
$H\!\parallel\!c$).
The Hall coefficient was described~\cite{Paschen} by an empirical
crossover function of the form $f(H,T) = A_2 -
(A_2-A_1)/[1+(H/H_0)^{p}]$; the crossover field scale $H_0(T)$ is
equivalent to an energy scale $T^*(H)$. We have analyzed the
magnetostriction data, as well as the existing magnetization data
($H\!\perp\!c$)~\cite{Tokiwa,Gegenwart_JPSJ}, with the same
crossover function.
Note that no
corresponding anomalies can be
resolved in the magnetization data for $H\!\parallel\!c$
~\cite{Gegenwart_JPSJ},
which is almost linear in $H$.
The solid curves in Fig.~1 and Fig.~2{\bf A} correspond to fits
of $\lambda_{[110]}$, $\tilde{M} \equiv M + \chi H$
and the Hall resistivity
$\rho_H$.
Fig.~2{\bf B} shows the three sets of
$H_0(T)$,
obtained from
such fits.
Their overlap represents a key conclusion of the present work;
it suggests that they define one energy scale
$T^*(H)$. This scale is seen to
be distinct from either the transition temperature ($T_N$) for the
magnetic ordering at $H<H_c$ or the scale ($T_{LFL}$) for the
establishment of the Landau Fermi liquid state at $H>H_c$. For all
three quantities, the width of the crossover extrapolates to zero at
$T=0$, implying that the differentials of the magnetostriction,
magnetization, and Hall resistivity have a jump in the
zero-temperature limit
(supporting online text).

The results raise the important question of the causal relationship
between the thermodynamic and electronic transport properties.
One might argue~\cite{HvL-RMP} that the Hall-effect evolution as a function
of the magnetic field~\cite{Paschen} is caused by the
Zeeman splitting of the Fermi surface induced by the magnetization
(and reflected in the magnetostriction).
However, the magnetization only
displays a smeared kink,
and the corresponding Fermi surface change would at most produce a
smeared kink in the Hall coefficient evolution; such a kink is too weak
compared to the smeared jump seen experimentally. Moreover, along
the $c-$axis, even such a smeared kink feature is absent in the
magnetization vs the magnetic field. Instead, it is more natural to
attribute the nonanalyticities in both the magnetostriction and
magnetization as thermodynamic manifestations of the large Fermi
surface jump caused by $f-$electron localization.

To explore this issue further, we have also studied the longitudinal
magnetoresistivity. Fig.~3 shows the electrical resistivity,
$\rho$, as a function of the magnetic field ($H\!\perp\!c$), at
various temperatures. The broadened step-like decrease, observed at
all temperatures, corresponds to the crossover observed in the other
properties. Indeed, as shown in Inset {\bf A}, the crossover fields
determined from the minima of the derivative $d\rho/dH$ ({\it cf.}
Inset {\bf B}) fall on the same $T^*(H)$ line determined from the
magnetostriction, magnetization, and Hall effect. In addition, Inset
{\bf B} shows that the width of the crossover decreases as
temperature is lowered. A detailed analysis shows that the crossover
width goes to zero in the zero-temperature limit
(supporting online text), implying a jump in
the residual resistivity across the magnetic QCP. This is in
accordance with the theoretical expectations~\cite{Coleman05,Si03}
associated with an $f$-electron localization transition.

Fig.~3, Inset {\bf A}, also shows the temperature scale as a
function of field, extracted from the peak in the $T$-dependence of
the differential susceptibility $\chi_{ac}=\partial M/\partial H$;
the latter, observed earlier\cite{Gegenwart}, necessarily
accompanies the smeared kink behavior in the isothermal $M$ vs $H$.
It is clearly seen that this scale too falls on the same $T^*(H)$
line.

Our results shed
%%new
light on the overall
phase diagram of this clean stoichiometric quantum critical
material. NMR measurements~\cite{Ishida}, while signaling the
dominance of AF fluctuations
in the quantum critical regime,
have also revealed enhanced ferromagnetic fluctuations:
the Korringa ratio, $S=1/T_1TK^2$, is small -- of the order of $0.1S_0$,
where $S_0$ is the corresponding ratio for non-interacting electrons.
Further evidence for enhanced ferromagnetic fluctuations has come
from magnetization measurements~\cite{Gegenwart}:
the Wilson ratio,
$R_W=\pi^2k_B^2/(\mu_0\mu_{eff}^2)\times\chi_0/\gamma_0$, with
$\mu_{eff}=1.4~\mu_B$/Yb
~\cite{Custers},
%% is strongly enhanced ($\sim 20$) for an extended
%%region of the phase diagram, including deep into the field-induced
%%paramagnetic Fermi-liquid phase; it further increases as the field
%%is reduced towards $H_c$.
is strongly enhanced for an extended region of the phase
diagram; it is already large ($\sim 20$) for magnetic fields of a few
Teslas, and further increases as the field is reduced towards $H_c$.
Therefore, it could be tempting to consider the ${\bf q \sim 0}$
magnetic fluctuation as the dominant critical fluctuation
\cite{HvL-RMP}, especially since a conventional ferromagnetic QCP
would yield a Gr\"uneisen exponent~\cite{Zhu-prl03} of $1/z\nu=2/3$,
close to what is observed in YRS~\cite{kuchler-prl03}.
This picture is problematic for a number of reasons, however. First,
neither 3D nor 2D ferromagnetic spin fluctuations can generate the
fractional exponent observed in the temperature dependence of the
uniform spin susceptibility \cite{Gegenwart}. Second, ferromagnetic
spin fluctuations would lead to a divergent
$1/T_1$ ($\sim 1/T^x$, with $x=1/3,1/2$ for 3D and 2D cases,
respectively), that is in contrast to the observation of $1/T_1 \sim
{\rm const.}$ when the NMR measurement field is extrapolated to the
quantum critical regime \cite{Ishidausr}.
Third,
%%since
as
ferromagnetic spin fluctuations are inefficient in
affecting
charge  transport, this picture contradicts the observation of a
nearly $H-$independent ratio $A/\chi^2$ that accompanies a strongly
$H-$dependent $A$ and $\chi$ \cite{Gegenwart}. Here, $A$ is the
coefficient of the $T^2$ component of the resistivity.

The data presented
here
show that the uniform  ($i.e.$, ${\bf q=0}$)
magnetization depends on
the same underlying physics as that for the charge transport.
Since the transport is dominated by large ${\bf q}$ fluctuations,
the results imply that the ${\bf q=0}$ magnetic fluctuations are a
part of overall fluctuations in an extended range of wavevector
scales.  It is then more natural to assume that the dynamical spin
susceptibility at different wavevectors obeys the same
form~\cite{Schroder,Si} as observed in another prototypical quantum
critical heavy fermion metal, CeCu$_{5.9}$Au$_{0.1}$~\cite{HvL-RMP}:
$\chi ({\bf q},T,\omega) \sim [\Theta_{\bf q} +
T^{\alpha}W(\omega/T)]^{-1}$. At the QCP, the Weiss field at the
antiferromagnetic wavevector (${\bf q}={\bf Q}$) vanishes:
$\Theta_{\bf Q}=0$. At the same time, and unlike for
CeCu$_{5.9}$Au$_{0.1}$, $\Theta_{\bf q=0}$ is very small in YRS.
Based on the saturation scale seen in the temperature dependence of
the uniform magnetic susceptibility~\cite{Gegenwart} and the NMR
Knight shift data~\cite{Ishida} near $H_c$, we estimate $\Theta_{\bf
q=0}$ to be of order 0.3~K
~\cite{note_cw}.
When ${\bf q}$ moves away from either
${\bf 0}$ or ${\bf Q}$, $\Theta_{\bf q}$ increases to the order of
the RKKY interaction or bare Kondo scale (about 25~K for
YRS~\cite{Trovarelli}). This is illustrated in Fig.~4. The
enhanced uniform magnetic susceptibility, the concomitant enhanced
Wilson ratio \cite{Gegenwart}, as well as the small $S \equiv
1/T_1TK^2$,
naturally follow from this picture.
Moreover, both $\chi_{\bf q=0}$ and $\chi_{{\bf q}={\bf Q}}$
scale similarly with $H$ and
the observation that $A/\chi^2$ is nearly $H$-independent
is in fact a manifestation of
an $H$-independent $A/\chi_{\bf Q}^2$.
All these lead to the conclusion that the origin of the $T^*$ line
lies in an electronic slowing down and, for YRS, the strong ${\bf
q}={\bf 0}$ fluctuations happen to be a consequence of the latter as
well.

We now turn to more detailed theoretical implications of our results.
Our measurements establish that the energy scale $T^*$
is associated with the equilibrium many-body spectrum
(which alone determines thermodynamics).
%%, and
Moreover,
this scale is distinct from  the Landau Fermi liquid
scale, $T_{LFL}$,
%%; moreover,
as physical quantities manifest rather different behavior across the
two scales (supporting online text). Finally,
both of these scales vanish at the QCP.
These findings
contradict the
conventional
order-parameter fluctuation theory in at least two respects. First,
the only low-energy scale in that theory is associated with the
magnetic slowing down which, for $H>H_c$, is
$T_{LFL}$
~\cite{Hertz,Millis,Moriya,Mathur}. Second, within that
theory, a sharp feature in thermodynamics and transport quantities
might arise near $T_N$ only.

Our results are instead consistent with magnetic quantum criticality
accompanied by the destruction of Kondo entanglement.
In the form of local quantum criticality~\cite{Coleman,Si},
a collapse of a large Fermi surface as $H$ decreases
leads
to an added energy scale characterizing an electronic slowing down
and, in addition,
yields a zero-temperature jump in the Hall coefficient and in the
field-differentials of the thermodynamic quantities.
An additional energy scale also
exists in the ``deconfined''
quantum criticality
%% scenario~\cite{Senthil}; however,
%%there is no evidence in our data for the fractionalized
%%Fermi liquid phase that arises in that theory (27).
scenario for insulating quantum magnets~\cite{Senthil},
as well as in its extension to itinerant electron
systems~\cite{Vojta,Senthil2} that are argued to be relevant
%%for
to
quantum critical heavy fermion metals.

%%To summarize,
%%we have provided the first
%%evidence that quantum criticality can contain low energy scales in
%%the equilibrium many-body spectrum that go beyond the one associated
%%with the slow fluctuations of the order parameter.
%%The existence of vanishing
%%multiple energy scales implies that, in contrast
%%to the
%%conventional wisdom, the theory of quantum criticality needs to incorporate
%%inherently quantum
%%degrees of freedom.

\bibliography{ref}
\bibliographystyle{Science}

%%%%%%%%%%%%%%%%%%%%%%%%%%% Acknowledgement%%%%%%%%%%%%%%%%%%%%%%%%%%%%%
% Following is a new environment, {scilastnote}, that's defined in the
% preamble and that allows authors to add a reference at the end of the
% list that's not signaled in the text; such references are used in
% *Science* for acknowledgments of funding, help, etc.

%%%%%%%%%%%%%%%%%%%%%%%%%%%%%%%%%%%%%%%%%%%%%%%%%%%%%%%%%%%%%%%%%%%%%%%%

% For your review copy (i.e., the file you initially send in for
% evaluation), you can use the {figure} environment and the
% \includegraphics command to stream your figures into the text, placing
% all figures at the end.  For the final, revised manuscript for
% acceptance and production, however, PostScript or other graphics
% should not be streamed into your compiled file.  Instead, set
% captions as simple paragraphs (with a \noindent tag), setting them
% off from the rest of the text with a \clearpage as shown  below, and
% submit figures as separate files according to the Art Department's
% instructions.

\clearpage

%%%%%%%%%%%%%%%%%%%%%%%Figure Captions%%%%%%%%%%%%%%%%%%%%%%%%%%%%%%%%%%%%%%%%
\begin{figure}[h!]
\caption{Magnetic field dependence of the magnetostriction of
YbRh$_2$Si$_2$.
The
%%open
symbols represent the linear coefficient
$\lambda_{[110]}=\partial \ln L /\partial H$ (where $L$ is the
sample length along the $[110]$ direction within the tetragonal $ab$
plane) vs $H$ at various temperatures.
Note, that $\lambda_{[110]}<0$ and that the data sets have been
shifted by different amounts vertically.
The sharp feature in the 0.02~K data corresponds to a discontinuity
in $\lambda$ (as is more clearly seen in the measured length vs $H$,
which shows a change in slope but does not contain any
discontinuity), demonstrating the continuous nature of the magnetic
transition at the critical field of 0.05~T.
Similar behavior is observed at various different temperatures
below 0.075~K, {\it e.g.} at 0.03~K, 0.04~K, 0.05~K and 0.06~K.
The solid lines for $T\geq 0.13$~K are
fits using the integral of the crossover function $f(H,T)$ (see
text) which reveal a characteristic field scale $H_0(T)$ along which
the magnetostriction shows a drastic change in slope.
} \label{FIG1}
\end{figure}

\begin{figure}[h!]
\caption{Energy scales in YbRh$_2$Si$_2$, determined from
thermodynamic, magnetic and transport measurements. ({\bf A}) the
field dependence ($H\perp c$)
of the magnetostriction $\lambda_{[110]}$, $\tilde{M}\equiv M + \chi
H$ (where $\chi =\partial M/\partial H$), and the Hall resistivity
$\rho_H$ for $H\parallel c$ (for the latter, the field values have
been multiplied by the factor 13.2 which is the anisotropy ratio of
the critical field parallel and perpendicular to the $c$-axis),
respectively, all at $T=0.5$~K; similar behavior is observed at
other temperatures. For $\lambda_{[110]}$($H\!\perp\!c$) the sample
has $\rho_0=0.5~\mu\Omega$cm and $H_c=0.05$~T, for $\tilde{M}$
($H\!\perp\!c$) the sample has $\rho_0=1.0~\mu\Omega$cm and
$H_c=0.06$~T, and for $\rho_H$($H\!\parallel\! c$) the sample has
$\rho_0=1.0~\mu\Omega$cm and $H_c=0.66$~T. The solid lines
correspond to fits using the integral of the crossover function
$f(H,T)$ (see text). Each data set has been normalized by its
initial slope. For clarity, the three data sets have been shifted by
different amounts vertically. $\rho_{H}-\rho_{H,a}$, where
$\rho_{H,a}$ is the anomalous Hall resistivity \cite{Paschen},
behaves similarly as $\rho_H$. We analyzed $\tilde{M}(H)$,
which is the field derivative of the magnetic free energy contribution
$(-M\times H)$; fitting
$M(H)$ leads to similar conclusions, though the quality of the fit
is somewhat poorer since $M$ vs $H$ is not as linear as $\tilde{M}$
at high fields (Fig.~S2).
~({\bf B}) the crossover field scale $H_0$ as determined from
magnetostriction, $\tilde{M}$ and Hall resistivity using the same
symbols as in {\bf A}. It is equivalent to an energy scale
$T^\star(H)$. The gray diamonds and triangles represent respectively
the N{\'e}el ordering temperature ($T_N$) and the crossover
temperature ($T_{LFL}$) below which the electrical resistivity has
the Fermi liquid form, $\rho=\rho_0+AT^2$, as determined from
measurements on a single crystal with $\rho_0=0.5~\mu\Omega$cm and
$H_c=0.05$~T. The solid and dotted lines are guides to the
eye;
for the latter, data
points outside the plotted field regime have also
been used. The horizontal error bars represent the fitting error
rather than the width of the crossover.} \label{FIG2}
\end{figure}

\begin{figure}[h!]
\caption{ Longitudinal magnetoresistivity of YbRh$_2$Si$_2$ as
$\rho$ vs $H$ ($H\!\perp\! c$) at various temperatures. The maxima
in the 0.03 and 0.07~K data indicate the boundary of the AF ordered
state ($T_N(H=0)=0.075$~K). The arrows mark the positions of
inflection points in $\rho(H)$. The inset {\bf A} displays the phase
diagram, where the gray shaded area represents the range of $H_0(T)$
values shown in Fig.~2{\bf B}. Open yellow triangles mark the
positions of the inflection points in the longitudinal electrical
resistivity. The smeared kink behavior in the isothermal $M$ vs $H$
corresponds to a peak in the $T$-dependence of $\chi$. The latter
has been observed \cite{Gegenwart}; the corresponding peak
temperature vs $H$ for a sample with $\rho_0=0.5~\mu\Omega$cm and
$H_c=0.05$~T is displayed by the dark blue diamonds and found to be
consistent with $H_0(T)$. The inset {\bf B} displays the derivative
$d\rho/dH$ vs $H$ at both $T=0.1$~K and 0.3~K. Arrows mark the
minima, corresponding to the inflection points in $\rho(H)$.}
\label{FIG3}
\end{figure}

\begin{figure}[h!]
\caption{ Sketch of the suggested {\bf q} dependence of the Weiss
temperature $\Theta_{\bf q}$, which enters the magnetic
susceptibility. As $H$ reaches $H_c$, it vanishes at the
antiferromagnetic wavevector ${\bf Q}$, as shown. In addition,
$\Theta_{\bf q}$ has a second minimum at ${\bf q}=0$. Ferromagnetic
fluctuations at ${\bf q=0}$ remain important as $\Theta_{\bf q=Q}$
goes to zero.} \label{FIG4}
\end{figure}

\newpage
%%%%%%%%%%%%%%%%%%%%%%%%%%%%%%%%%%%%%%%%%%%%%%%%%%%%%%%%%%%%%%%%%%%%%%%%

%%\vspace{0.5cm}

\begin{scilastnote}
\item We would like to thank
P. Coleman, J. Custers, Z. Fisk, S. Friedemann, K. Ishida, S.
Kirchner, D. Natelson, N. Oeschler, and S. Wirth
for useful discussions. This
work has been supported by the Fonds der Chemischen Industrie, NSF
Grant No. DMR-0424125 and the Robert A. Welch Foundation.
\end{scilastnote}

\vskip 0.5 cm

\noindent{\bf Supporting Online Material}

%\noindent{www.sciencemag.org}

\noindent{Supporting online text}

\noindent{Figs.~S1,S2}

\newpage
\begin{figure}[h!]
\centerline{\includegraphics[width=0.8\linewidth]{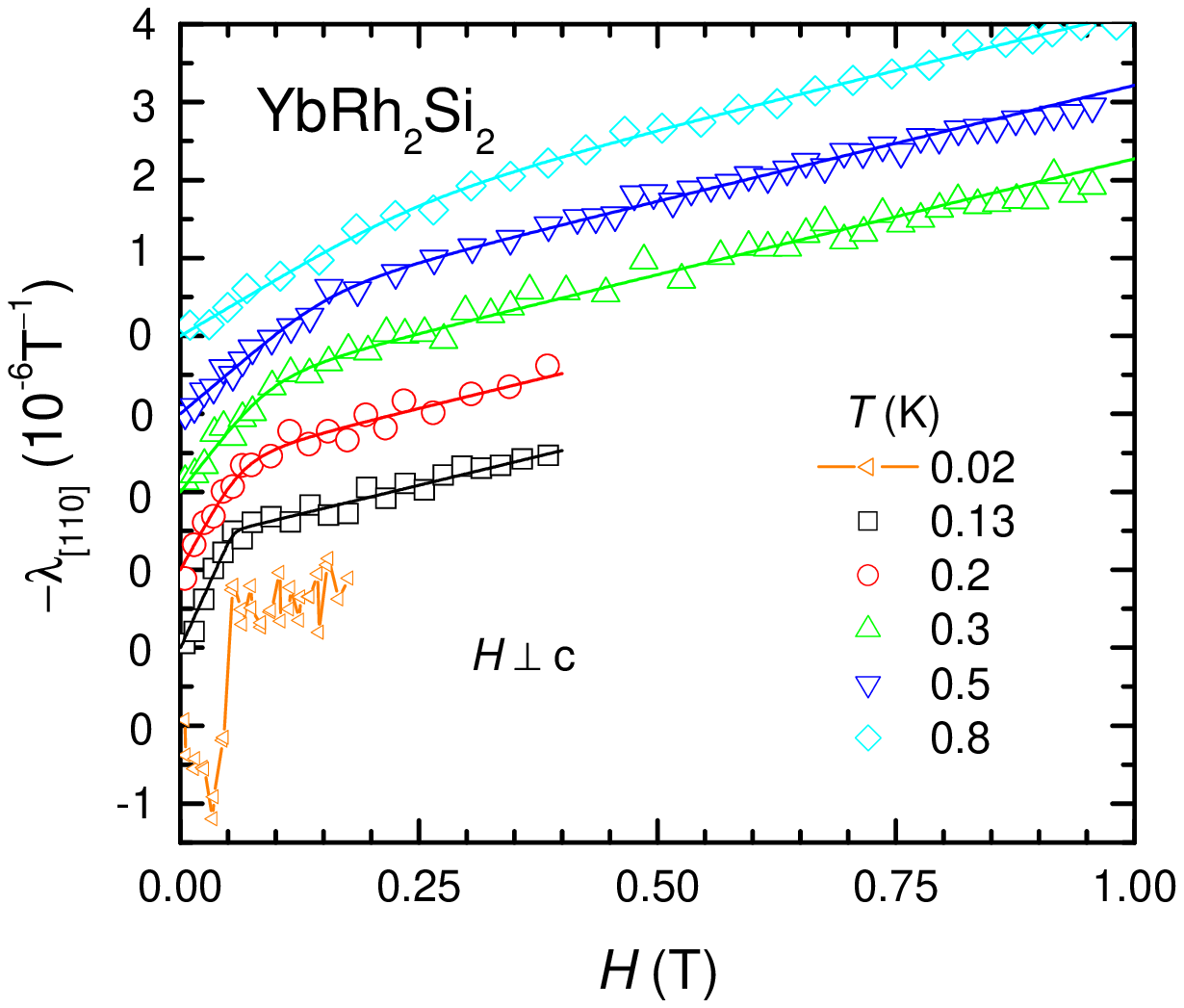}}
\centerline{} \centerline{} \centerline{\Large Figure~1}
\end{figure}

%\newpage

\begin{figure}[h!]
\centerline{\includegraphics[width=0.8\linewidth]{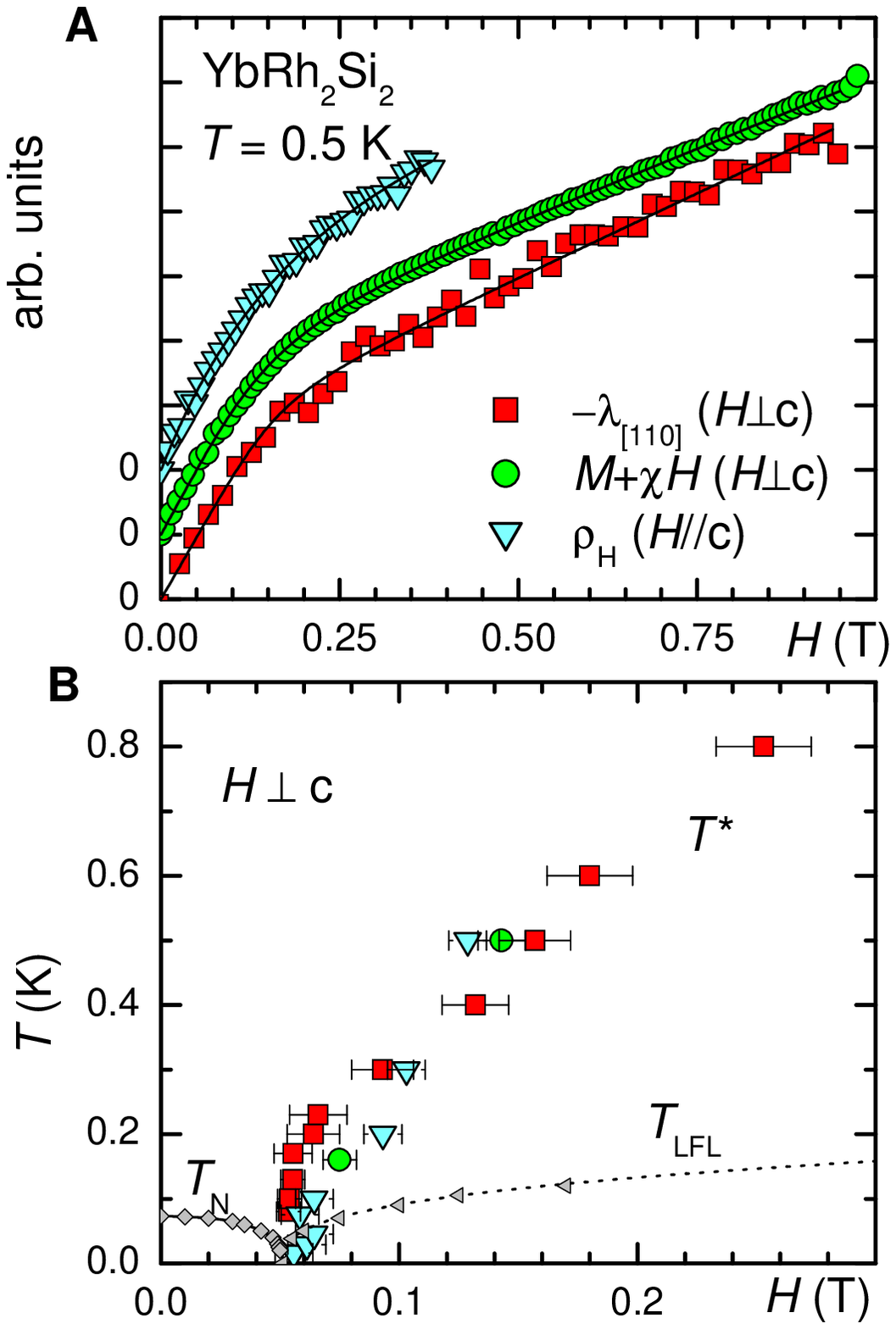}}
\centerline{} \centerline{} \centerline{\Large Figure~2}
\end{figure}

%\newpage

\begin{figure}[h!]
\centerline{\includegraphics[width=0.8\linewidth]{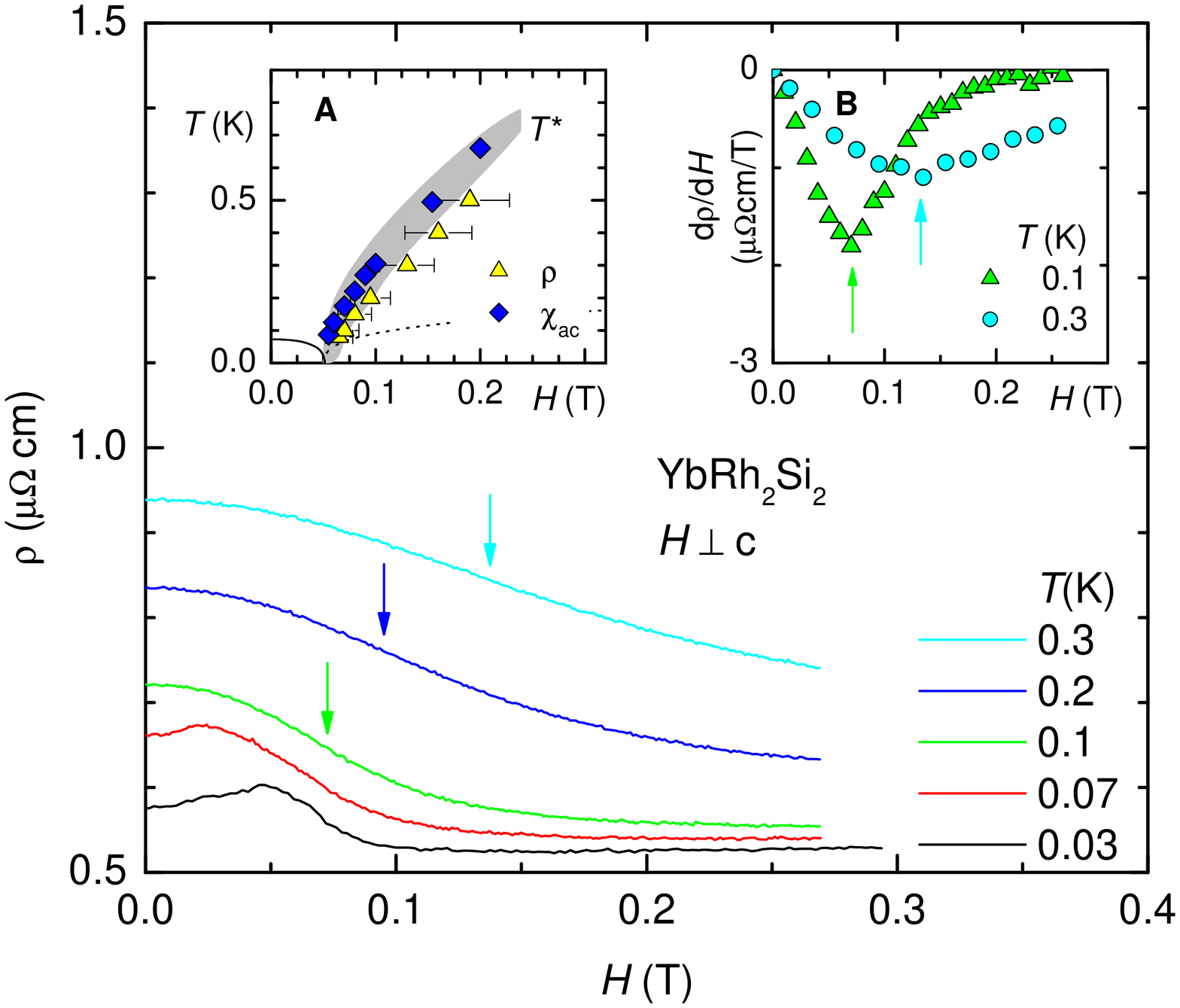}}
\centerline{} \centerline{} \centerline{\Large Figure~3}
\end{figure}

%\newpage

\begin{figure}[h!]
\centerline{\includegraphics[width=0.8\linewidth]{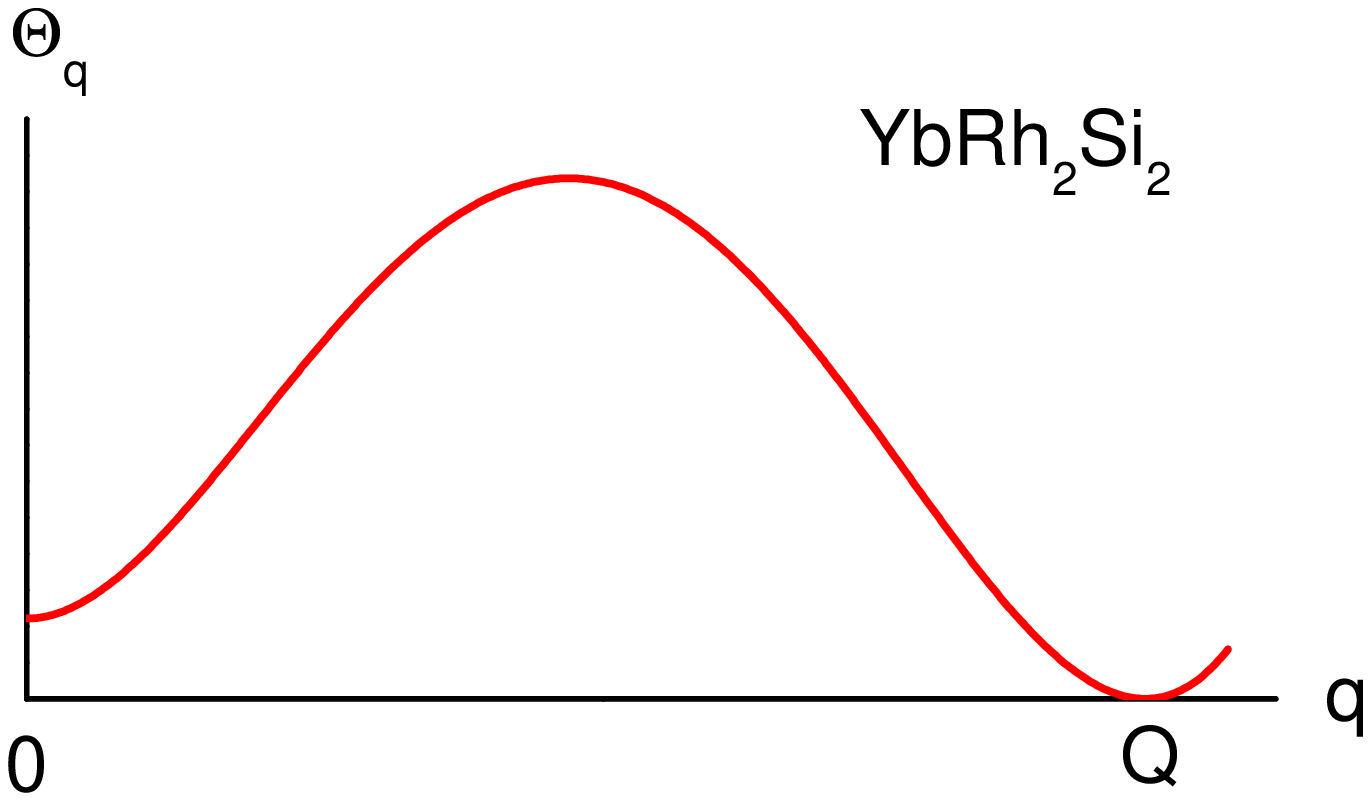}}
\centerline{}
\centerline{}
\centerline{\Large Figure~4}
\end{figure}

\cleardoublepage
\noindent {\bf SOM 1.~Relationship between $T^*$
and $T_{LFL}$}

The two scales, $T^*$ and $T_{LFL}$, are distinct due to the rather
different behavior that physical properties manifest across these
scales.

Across the $T^*$ line, as a function of the magnetic field ($H$),
the behaviors of isothermal physical quantities are governed by a
{\it zero-temperature jump} that is smeared by temperature. The
physical quantities include $d\lambda/dH$,
$\tilde{\chi}=d\tilde{M}/dH$, $d \rho_H /d H$, and $\rho$. The FWHM,
defined in the main text, for each of these quantities is plotted as
a function of temperature in Fig.~S1{\bf A}. Each FWHM extrapolates
to zero in the $T=0$ limit, signaling a jump in the zero-temperature
limit. For each quantity, the ratio $A_2/A_1$ (where $A_2$ and
$A_1$, defined in the main text, are respectively the low-field and
high field values) is plotted as a function of temperature in
Fig.~S1{\bf B}. It extrapolates to a value significantly below $1$
in the $T=0$ limit, implying that the extrapolated jump at zero
temperature is non-zero. The existence of this zero-temperature jump
implies that none of the four quantities obeys a scaling in terms of
$T/T^*$.

The $T_{\rm LFL}$ line, on the other hand, represents a very
different crossover. This line was already extensively discussed, %%
based upon the specific heat and resistivity data, in Ref.~22 of the
main text. For resistivity, this scale appears in the temperature
dependent component. Indeed, $d\rho/dT$ satisfies a scaling in terms
of $T/T_{LFL}$:
\begin{eqnarray}
\frac{d\rho}{dT} = \phi \left ( \frac{T}{T_{LFL}(H)} \right  )
\label{scaling}
\end{eqnarray}

Alternatively, the contrast between the $T_{\rm LFL}$ line and the
$T^*$ line can be seen by integrating the scaling equation
(\ref{scaling}), yielding a $\rho(H,T)-\rho(H,0)$ vs. $H$ that
always has a smooth crossover across the $T_{LFL}$ line. This is in
contrast to the jump associated with the $T^*$ line discussed
earlier.

\vskip 0.5 cm

\noindent {\bf SOM 2.~$M$ vs. $H$}

As shown in Fig.~S2, the magnetization itself displays a similar
crossover as $\tilde{M} \equiv M + \chi H$ does. Compared to the
$\tilde{M}$ case, its linear $H$ dependence in the measured high
field regime is somewhat less robust, making the fit by the fitting
function $f(H,T)$ to be of a slightly lower quality.

%\newpage
\setcounter{figure}{0}
\renewcommand\thefigure{S\arabic{figure}}
\begin{figure}[h!]
\centerline{\includegraphics[width=0.5\linewidth]{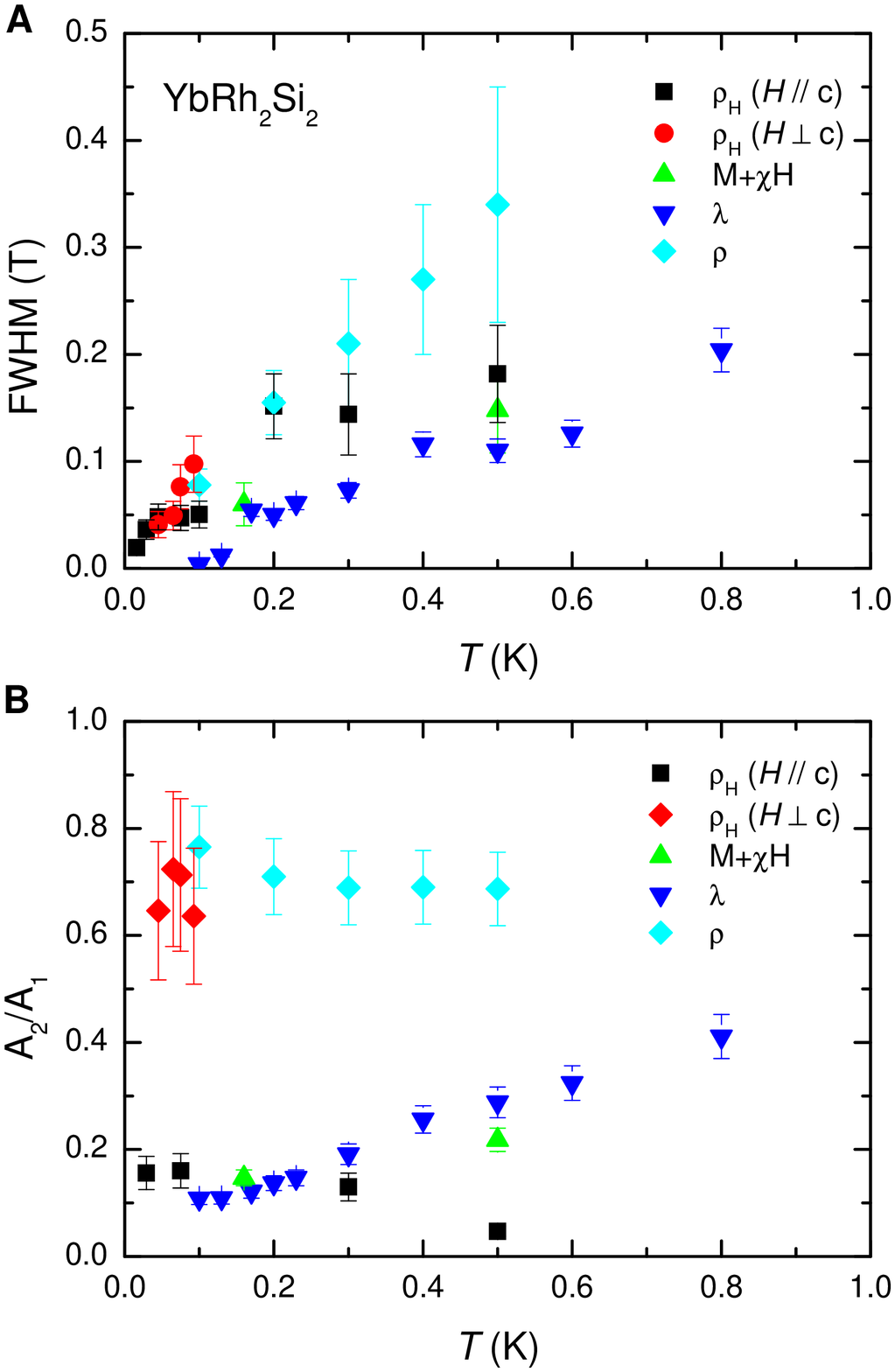}}
\caption{({\bf A}) Temperature dependence of the full-width at
half-maximum (FWHM) derived from fitting transport and thermodynamic
measurements across $T^\star(H)$ by the crossover function $f(H,T) =
A_2 - (A_2-A_1)/[1+(H/H_0)^{p}]$. For the Hall resistivity $\rho_H$
for $H\parallel c$, the values have been divided by 13.2 which is
the anisotropy ratio of the critical field parallel and
perpendicular to the $c$-axis. ({\bf B}) Temperature dependence of
the ratio $A_2/A_1$.} \label{FIG1}
\end{figure}

\begin{figure}[h!]
\centerline{\includegraphics[width=0.5\linewidth]{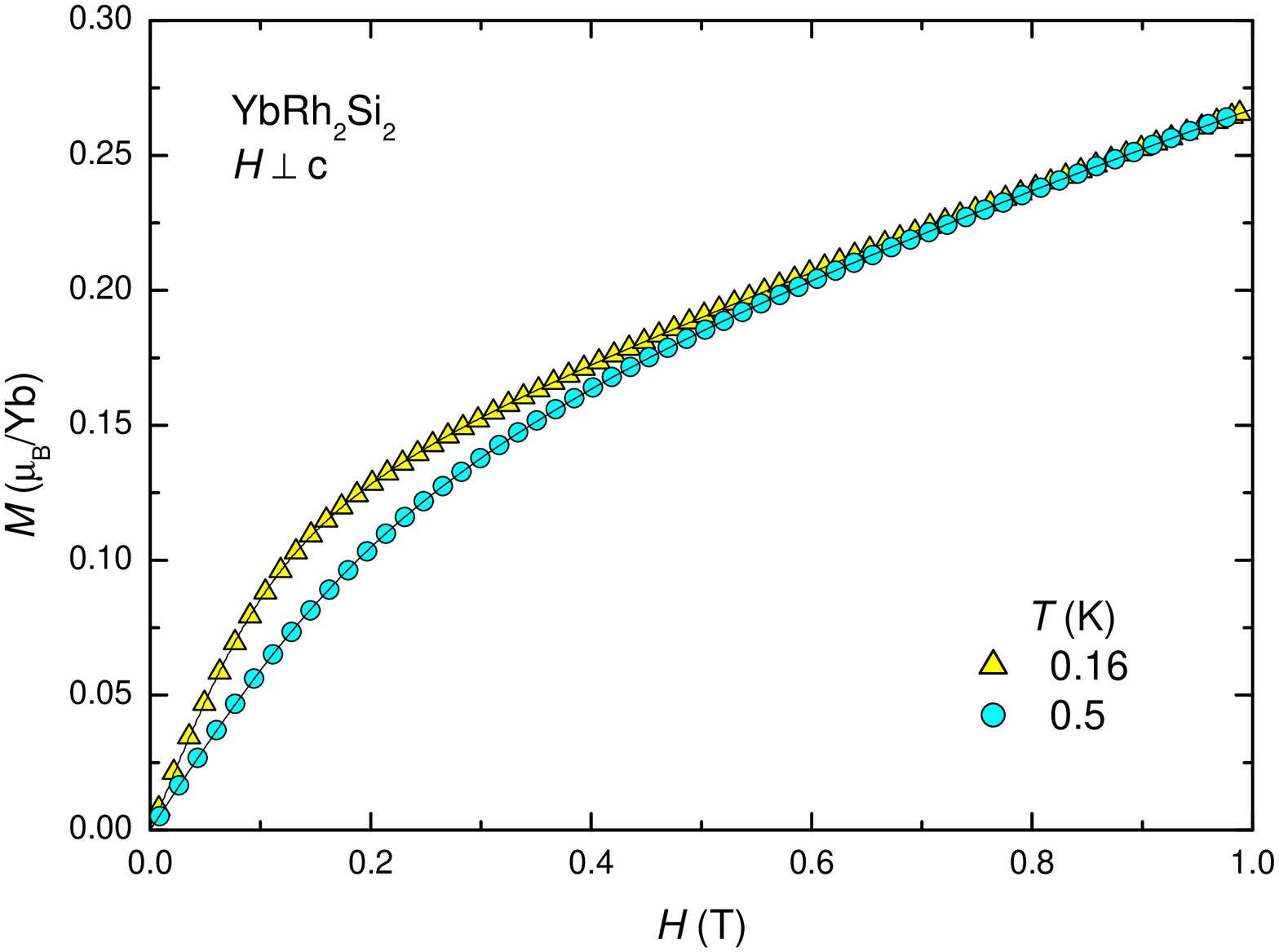}}
%%\centerline{\includegraphics[width=0.5\linewidth]{FIG4.eps}}
\caption{Magnetic field dependence of the Magnetization $M(H)$ of
YbRh$_2$Si$_2$ at 0.16 K (triangles) and 0.5 K (circles). Lines are
fits according to the crossover function with $B_0=0.11$~T (0.16~K)
and $B_0=0.2$~T (0.5~K).} \label{FIG2}
\end{figure}

\end{document}